\documentclass{iaus}
\usepackage{graphicx}

\title[Cl and S in nearby nebulae] 
{Chlorine and Sulfur in Nearby Planetary Nebulae and H~II Regions}

\author[M\'onica Rodr\'iguez \& Gloria Delgado-Inglada]   
{M\'onica Rodr\'iguez
 \and Gloria Delgado-Inglada}

\affiliation{Instituto Nacional de Astrof\'isica,
  \'Optica y Electr\'onica (INAOE), Apdo Postal 51 y 216, 72000 Puebla,
  Mexico, email: {\tt mrodri@inaoep.mx, gloria@inaoep.mx}}

\pubyear{2011}
\volume{283}  
\pagerange{1--2}
\jname{Planetary Nebulae: an Eye to the Future}
\editors{A. Manchado, L. Stanghellini \& D. Schoenberner, eds.}
\begin{document}

\maketitle

\begin{abstract}
We derive the chlorine abundances in a sample of nearby planetary nebulae (PNe)
and H~II regions that have some of the best available spectra. We use a nearly
homogeneous procedure to derive the abundance in each object and find that the
Cl/H abundance ratio shows similar values in H~II regions and PNe. This supports
our previous interpretation that the underabundance we found for oxygen in the
H~II regions is due to the depletion of their oxygen atoms into organic
refractory dust components. For other elements, the bias introduced by ionization
correction factors in their derived abundances can be very important, as we
illustrate here for sulfur using photoionization models. Even for low-ionization
PNe, the derived sulfur abundances can be lower than the real ones by up to 0.3
dex, and the differences found with the abundances derived for H~II regions that
have similar S/H can reach 0.4 dex.
\end{abstract}


Recently, we derived in a homogeneous way the oxygen abundances for five H~II
regions and eight planetary nebulae (PNe) of the solar neighborhood (closer than
2 kpc), using available spectra of high quality (\cite[Rodr\'iguez \&
Delgado-Inglada 2011]{rod11}). We used collisionally excited lines and
recombination lines of oxygen, finding  that in both cases the abundances derived
for the PNe are $\sim0.2$~dex above those calculated  for the H~II regions. We
compared the resulting abundances with those found for the Sun, B stars, and the
diffuse interstellar medium (ISM). A good agreement can be reached for the
results implied by collisionally excited lines if the H~II regions have about a
quarter of their oxygen atoms deposited in an organic refractory dust component.
This dust component was previously  proposed to explain the pattern of oxygen
depletion in the diffuse and dense ISM (\cite[Jenkins 2009]{jen09}; \cite[Whittet
2010]{whi10}). Oxygen is the element for which the derived abundances are more
reliable, but the abundances of the other elements could provide further evidence
on this issue. Here we present results for chlorine and sulfur. 

\begin{figure}[tb]
 \vspace*{-0.3 cm}
\begin{center}
 \includegraphics[width=3.2in]{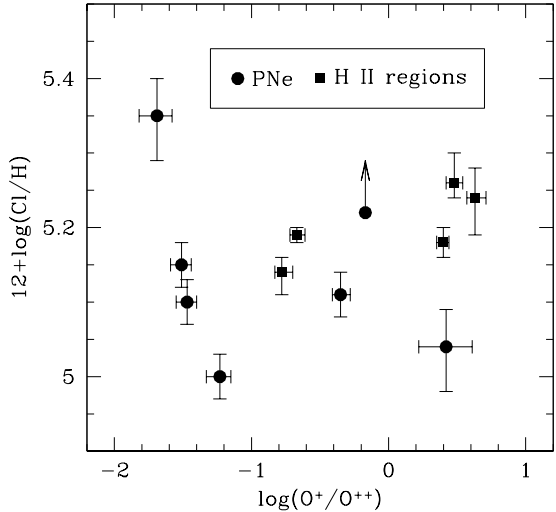} 
 \vspace*{-0.5 cm}
 \caption{Chlorine abundances for the sample nebulae as a function of
 $\mbox{O}^+/\mbox{O}^{++}$.}
   \label{fig1}
\end{center}
\end{figure}

Figure~\ref{fig1} shows the chlorine abundances calculated using the physical
conditions derived previously (see \cite[Rodr\'iguez \& Delgado-Inglada
2011]{rod11}) and the intensities of [Cl~II] $\lambda$9124, [Cl~III]
$\lambda\lambda$5517, 5537, and [Cl~IV] $\lambda$8046 or [Cl~IV] $\lambda$7530.
[Cl~II] lines were not available for the PNe with $\mbox{O}^+/\mbox{O}^{++}<1$,
but the Cl$^+$ concentration is small in high ionization objects.
On the other hand, [Cl~IV] $\lambda$8046 (used for the H~II regions) was not in
the wavelength range observed for the PNe, whereas [Cl~IV] $\lambda$7530 (used
for the PNe), which arises from the same upper level, is weaker and probably
blended with a C~II line in the H~II regions. The [Cl~IV] lines were outside of
the observed ranges for two PNe (one has a lower limit in the figure; the other
one falls below the plotted range), and unobserved in the low-ionization objects,
where the contribution of Cl$^{3+}$ should be negligible. Chlorine is not
expected to be further ionized in our sample objects. Because of the small
differences in the procedures used to derive Cl/H in the different objects, the
comparison is less reliable than in the case of oxygen, but the results in the
figure suggest that nearby H~II regions and PNe have similar abundances of
chlorine. This supports the idea described above that oxygen is more depleted in
H~II regions than in PNe.

Besides oxygen and chlorine, the abundances of sulfur, neon, and argon are not
expected to be modified during the evolution of the PNe progenitor stars and can
also be derived using optical spectra. However, in order to derive the abundances
of these elements we need to study the bias introduced by the ionization
correction factors (ICFs), which is likely to be different in H~II regions and
PNe. We use sulfur to illustrate this issue, since the same ICF is generally
applied for this element in H~II regions and PNe. The ICF was originally derived
by \cite[Stasi\'nska (1978)]{sta78} for H~II regions, and was later adopted by
\cite[Kingsburgh \& Barlow (1994)]{kin94} for PNe. We calculated a series of
photoionization models with Cloudy (\cite[Ferland et al.\ 1998]{fer98}), ionized
by either O-type stars (\cite[Pauldrach et al.\ 2001]{pau01}) or evolved stars
(\cite[Rauch 2003]{rau03}). The models have  metallicities similar to those in
our sample objects, electron densities around 2000 cm$^{-3}$, and either
plane-parallel geometry (for the H~II regions) or spherical geometry (for the
PNe). We analyzed the predicted spectra to derive physical conditions and ionic
and total abundances following the same procedure we would use for real objects.
We compared the derived sulfur abundances with the input ones, finding that for
the model H~II regions the derived abundances are larger by up to $\sim0.1$ dex,
whereas for the model low-ionization PNe, like those in our sample, the derived
abundances can be higher by $\sim0.05$ dex or lower by $\sim0.3$ dex. Hence,
extreme care will be needed to achieve meaningful comparisons of the sulfur
abundances.

\end{document}